\documentclass[useAMS,usenatbib]{mn2e}
\usepackage{graphicx}
\usepackage{natbib}
\usepackage{amsmath}
\usepackage{amssymb}
\usepackage{graphicx}

\title[Supercritical Starless Cores]{Dynamics and Depletion in Thermally Supercritical Starless Cores}
\author[Keto and Caselli]{Eric Keto$^{1}$\thanks{E-mail:
keto@cfa.harvard.edu (EK); \hfill\break
p.caselli@leeds.ac.uk (PC)} and Paola
Caselli$^{2}$
\\
$^{1}$Harvard-Smithsonian Center for Astrophysics, 160 Garden St, Cambridge, MA 02420, USA \\
$^{2}$School of Physics and Astronomy, University of Leeds, Leeds LS2 9JT, UK
}
\begin{document}

\date{July 7, 2009}

%\pagerange{\pageref{firstpage}--\pageref{lastpage}} \pubyear{2009}

\maketitle

\label{firstpage}

\begin{abstract}
In previous studies we identified two classes of starless cores, thermally
subcritical and supercritical, distinguished by different dynamical behavior and
internal structure. Here we study the evolution of the dynamically-unstable,
thermally-supercritical cores by means of a numerical hydrodynamic
simulation that includes radiative equilibrium and simple molecular
chemistry. From an initial state as an unstable
Bonnor-Ebert (BE) sphere, a contracting
core evolves toward the configuration of a singular isothermal sphere (SIS)
by inside-out collapse. 
We follow the gas temperature and abundance of CO
during the contraction. 
The temperature is predominantly
determined by radiative equilibrium, but in the rapidly contracting 
center of the core, compressive
heating raises the gas temperature
by a few degrees over its value in static equilibrium. The time scale for the
equilibration of CO depends on the gas density and is everywhere shorter
than the dynamical timescale.
The result is that the dynamics do not much affect the abundance of CO 
which is always close to that of a static sphere of the same density profile,
and CO cannot be used as a chemical clock in starless cores.
We use our non-LTE radiative
transfer code MOLLIE to predict observable CO and N$_2$H$^+$
line spectra, including the non-LTE hyperfine ratios of N$_2$H$^+$, during the
contraction. These are
compared against observations of the starless
core L1544. The comparison indicates that the dust in L1544 has an
opacity consistent with ice-covered rather than bare grains, the cosmic ray
ionization rate is about $1\times 10^{-17}$ s$^{-1}$, and the density 
structure of L1544 is approximately
that of a Bonnor-Ebert sphere with a maximum central 
density of $2\times 10^7$ cm$^{-3}$, equivalent to an average density of
$3\times 10^6$ cm$^{-3}$
within a radius of 500 AU.  The observed CO line widths 
and intensities are reproduced if the CO desorption rate
is about 30 times higher than the rate expected from cosmic-ray 
strikes alone, indicating that
other desorption processes are also active.
\end{abstract}

\section{Introduction}

The starless cores are well described as small, dense, self-gravitating clouds supported
largely but not entirely by thermal pressure. They are significant in the interstellar medium
as the future birthplaces of stars \citep[reviews:][]{DiFrancesco2007,BerginTafalla2007}.
In previous papers \citep{KetoField2005, KetoCaselli2008} we developed a simple model
to describe the temperature, density, chemistry, and dynamical evolution of
the starless cores.
Those studies suggest that the starless cores can be conceptually divided into two classes,
thermally subcritical and thermally supercritical depending on whether
their central densities are very roughly less or greater than
a few $10^5$ cm$^{-3}$ respectively. This density is significant as
the approximate value of
two critical densities in the physics of starless cores.
First, at this density gas cooling by collisional
coupling with dust is about as efficient as cooling
by molecular line radiation.
The gas temperature in the center of the supercritical cores
is therefore significantly lower than in their envelopes
and lower than found anywhere in the subcritical cores.
Second, at this density, cores of a few M$_\odot$
are at critical dynamical stability with respect
to gravitational collapse. The thermally subcritical cores are stable against
gravitational collapse and their dynamics may be dominated by
oscillations \citep{Lada2003,Aguti2007}.
In contrast, the supercritical
cores are unstable with predominantly inward velocities
\citep{Williams1999, Caselli2002, Keto2004, Sohn2007, Schnee2007}

In previous papers \citep{Keto2006, Broderick2007, Broderick2008} we
modeled the internal oscillations (sound waves)
in thermally subcritical cores and computed some observable
molecular spectral line profiles during the oscillations. In this paper,
we model the dynamical and chemical
evolution of a thermally supercritical core and
compute  observable molecular spectral line
emission during the early stages of gravitational contraction
leading up to free-fall collapse.

Our model is a one-dimensional numerical
hydrodynamics code \citep{KetoField2005}
that includes the radiative equilibrium of dust,
gas cooling by both molecular lines and
collisional coupling with the dust, and a simplified
molecular chemistry \citep{KetoCaselli2008}.
To compare
the model with observations we use our numerical code for non-LTE radiative transfer
MOLLIE to predict
the observable molecular spectral line emission
\citep{Keto1990,Keto2004}. Previously 
we modeled overlapping hyperfine emission by assuming
that the hyperfine components are in statistical equilibrium. Here we use
an updated algorithm that does not make this assumption and better
reproduces the non-LTE hyperfine line ratios (excitation anomalies) that
are seen in N$_2$H$^+$ spectra from
many of the starless cores \citep{Caselli1995, Daniel2007}.
We compare these predictions to previous observations of one
specific core, L1544 whose properties place it in the thermally
supercritical class.

There has been some considerable research into the topic of how
gas clouds contract to form protostars including some classic
papers in the field \citep{Henyey1955, Hayashi1961, Larson1969, Shu1977}.
More recent studies also address the chemical evolution of the
starless cores using a variety of models for the dynamics of gravitationally
contracting cores.

\citet{Aikawa2001, Aikawa2003} followed the chemical evolution of a
gravitationally collapsing core assuming
the Larson-Penston (LP) approximation for the
densities and velocities. In a subsequent paper,
\citet{Aikawa2005} improved on this model using a numerical
hydrodynamic code \citep{Ogino1999} to follow the
collapse of an isothermal Bonnor-Ebert (BE) sphere.
\citet{Evans2001} and \citet{GalliWalsmleyGoncalves2002}
showed that in the case of
static BE spheres there was little
difference between the structure of an isothermal
BE sphere and one in radiative equilibrium.
Our numerical hydrodynamic simulation
that allows for radiative equilibrium
during collapse also confirms 
that the isothermal approximation used
in \citet{Aikawa2005} is quite
adequate for dynamic as well as static BE spheres (our \S \ref{insideout}).
However, if we go on to predict molecular
line strengths from the model cores, we need to consider
temperature variations. Our  \S \ref{lines} discusses the spectral line
modeling.

\citet{RawlingsYates2001} used the similarity solution of the collapse of a
singular isothermal sphere (SIS) \citep{Shu1977} in their study of chemical
evolution. \citet{Tsamis2008}  combined this model for the dynamics with
a model for the internal
temperature structure of a BE sphere from \citet{Evans2001}.
\citet{LeeBerginEvans2004} approximated the contraction of a BE sphere
as a progression of static spheres of increasing central density.
Our hydrodynamic simulation suggests that
this series of static equilibria is a good approximation during
the subsonic phase of contraction, and that
the SIS model is a good approximation 
at the endpoint of this phase (our \S \ref{contraction}).

\citet{VanWeeren2009} and \citet{Brinch2008} modeled the chemical
evolution during the formation of a 2D rotating accretion disk
and protostar. Our spherical model applies to earlier evolutionary
times while the contracting core still maintains its BE structure.

\citet{Li2002} and \citet{Shematovich2003} followed the chemical evolution
during the contraction of cores that are supported by a spherical approximation
of magnetic pressure. The contraction is controlled by  leakage
of the magnetic field out of the core by ambipolar diffusion. In the spherical
approximation, this is modeled by a diffusion equation in Lagrangian
coordinates \citep[equation 3 of ][]{Li2002}. One difficulty with
assessing the applicability of spherical models of magnetic collapse is that
the approximation of spherical symmetry results in a density structure
that, similar to the SIS and BE spheres, can be approximated by a
polytrope with an index that depends on the effective equation of
state. Observationally, it is difficult to distinguish polytropes of
different index because their density structures are most
different from one another at small and large radii where the observations have
difficulties with angular resolution and sensitivity (low density gas)
respectively.

In contrast, there is strong observational motivation for the model of
thermal support. Masses, densities, and
temperatures estimated in surveys of starless cores suggest that thermal energy
provides about 75\% of the internal
energy required for equilibrium \citep{DickmanClemens1983, MyersBenson1983, Lada2008}.
The remaining 25\% could be
in magnetic energy or wave energy or a combination of both.
While the 25\% non-thermal energy is important for the
dynamical stability and evolution
of the core, the structure of the core is determined
largely by the dominant thermal energy.

Our study is different from previous studies in that we consider the
evolution of a BE sphere rather than an accretion disk plus protostar,
or an SIS sphere,
or a magnetically supported sphere.
Our simulations is the first that follows the dynamics of the 
contraction simultaneously and self-consistently
with radiative equilibrium and a simple model for the chemistry that allows
for changes in the molecular line cooling as the molecular abundances in
the gas phase change with time.  

We first
discuss the dynamics of the gravitational contraction of a BE sphere.
Our code is best suited for subsonic velocities and we confine our
discussion to the early phase of evolution. We then discuss how the
gas phase abundance of CO evolves during contraction. We
predict how the molecular line strengths should vary across the
model from center to edge at different times during the contraction,
and we compare this prediction with ratios of line strengths
previously observed in L1544. We intend the comparison in a general sense
only and not in detail. For example, the model is spherically
symmetric while both the observed morphology
\citep{WardThompson1999,Bacmann2000} and observed spectra
\citep{Williams1999,Caselli2002,Williams2006} indicate that both the density and
velocity structure of L1544 are more complex than can be reproduced
by a purely spherical model.

\section{The gravitational contraction of an unstable Bonnor-Ebert sphere}\label{contraction}

We model a core of 10 M$_\odot$ that begins its evolution
with an initial density of $2 \times 10^4$ cm$^{-3}$ in a state
of radiative and dynamical equilibrium. The mass and density place the
core in the thermally supercritical class
\citep[figure 14 of][]{KetoCaselli2008} so that the dynamical equilibrium is
unstable.
Based on the results of
\citet{KetoCaselli2008} we use our higher dust opacities that are
equal to four times the standard dust opacities of \citet{OssenkopfHenning1994},
and we use a rate of cosmic ray ionization of
$1.3\times 10^{-17}$ s$^{-1}$ referred to as the "low" rate in
\citet{KetoCaselli2008}.  In that investigation, this combination
was found to result in gas temperatures that match those suggested
by the observations of \citet{Crapsi2007} at their suggested
central density of $2\times 10^6$ cm$^{-3}$.

Figures \ref{fig:densities} and \ref{fig:velocities} show the density and velocity profiles during
contraction as the central density
increases from its initial value to $2 \times 10^{8}$ cm$^{-3}$
over a time period of 1 Myr.
The shapes of the density and velocity profiles that are generated
by the numerical simulation can be understood in terms of relatively
simple theory. We begin with the density profile.

\subsection{The density profile}

The density profiles in figure \ref{fig:densities} show an outer region where the
density scales as $r^{-2}$ and an inner region where the
density is nearly constant. As the evolving core contracts,
the density profile maintains this
shape even as the central density increases
\citep{Larson1969, Hunter1977, FC1993, Ogino1999, Kandori2005}.
Thus we should expect to observe this
characteristic shape in almost all starless cores, and this
is generally confirmed by observations.
\citet{WardThompson1994}, \citet{Andre1996},
\citet{WardThompson1999},
and \citet{Bacmann2000}
suggested that the density profiles derived from
observations of dust emission could be well
matched by two power laws, one shallow and one
steep.
This approximation is also suggested by the numerical studies
of the contraction of a BE sphere
cited above.
 \citet[][\S 4.2]{Tafalla2002}
suggested that the density profile can be better fit by,
\begin{equation}\label{eq:plummer}
\rho(r) = {{\rho_c} \over {1 + (r/r_f)^\alpha}}
\end{equation}
where $r_f$
is the radius of the inner region. This
equation along with $r_f$, which we define
below, provides a convenient approximation
for future use.

\subsubsection{The flat inner region}

In the center of a BE sphere,
where the
sound-crossing time is less than
the free-fall time, density perturbations
are rapidly smoothed by pressure waves.
Also the self-gravity in the center
is relatively weak so that the gas is essentially
confined by the pressure of the overlying gas.
Both these effects act to maintain constant density
around the center.

Previously, the size of the region with a
flat density profile, $r_f$, was determined
empirically for individual cores from their observations.
We can also determine this radius theoretically. At $r_f$
the ratio of the sound-crossing
and free-fall times ought to be one.
Therefore, $r_f$ is
the product of the sound speed, $a$, and the free-fall time at the
central density, $\rho_c$,
\begin{equation}\label{eq:criticalRadius}
r_f = {{a} \over { ( 32\rho_c G/ 3\pi )^{1/2} }}.
\end{equation}
This provides a better approximation in equation 1 than
the standard
scale length of a BE sphere, $[kT/(4\pi G\rho_c)]^{1/2}$ \citep{Bonnor1956}
and is also useful
in suggesting the physics behind the behavior of a contracting
BE sphere. The radius, $r_f$ is equal to the Jeans length
for sound speed, $a$, and density, $\rho_c$.

\begin{figure}%[t]
\includegraphics[width=3.25in]{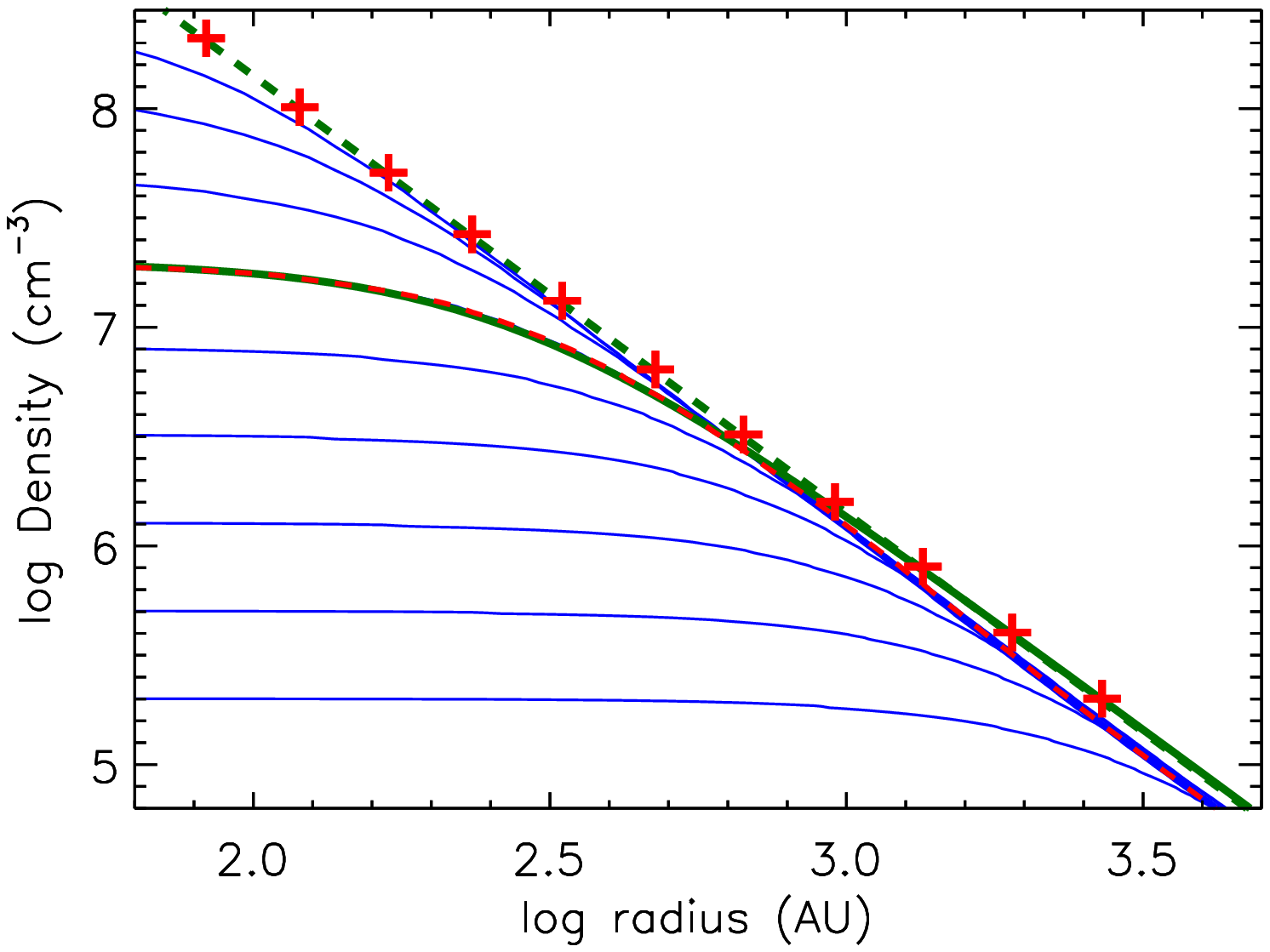}
\caption{The progression of the density profile of a thermally supercritical
starless core during subsonic contraction. The profile
with a central density of $2\times 10^7$ cm$^{-3}$ (the central density
of L1544) is indicated by the dashed
red line. The solid green line shows how well the approximation
of equation \ref{eq:plummer} fits this density profile even though the
approximation assumes isothermal gas and the numerical calculation
allows for temperature variations due to radiative equilibrium (figure \ref{fig:structure123}).
Density profiles at earlier and later times are shown in blue below and above
this line. The evolution covers 1 Myr. For each profile,
a cross marks the characteristic radius, $r_f$ (equation \ref{eq:criticalRadius})
that separates the inner region with approximately constant density, and the outer
region with density scaling as $r^{-2}$. A slope of $-2$ is shown as the dashed green line.
Over time, as the central density increases, the characteristic radius moves inward
and the BE sphere resembles more and more a singular isothermal sphere
with $\rho\sim r^{-2}$ everywhere.
}
\label{fig:densities}
\end{figure}

\begin{figure}%[t]
\includegraphics[width=3.25in]{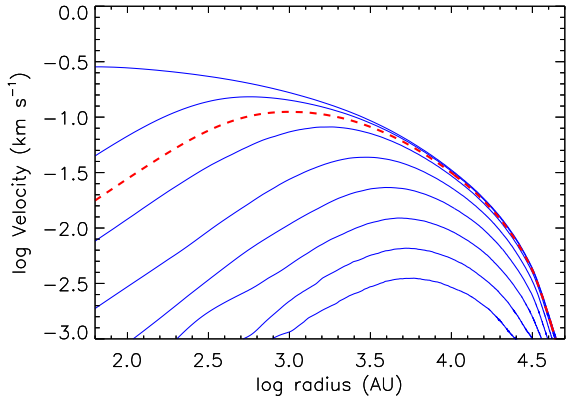}\break
\caption{The progression of the velocity profile of a thermally supercritical core
during subsonic contraction. The profiles have a characteristic $\Lambda$-shape
with the maximum inward velocity near the characteristic radius, $r_f$ (equation
\ref{eq:criticalRadius}). The profile at the
time that the central density is $2\times 10^7$ cm$^{-3}$ (the central density
of L1544) is indicated by the dashed
red line.
}
\label{fig:velocities}
\end{figure}

Figure \ref{fig:densities} compares the density profile computed
by equations \ref{eq:plummer} and \ref{eq:criticalRadius}, assuming
a gas temperature of 10 K and a central density of
$2 \times 10^7$ cm$^{-3}$ (green curve),   against
the density profile computed by the numerical hydrodynamic simulation.
The comparison suggests that equations \ref{eq:plummer} and
\ref{eq:criticalRadius} provide a good, simple approximation
to the density profile of a BE sphere.
(A more accurate approximation
is described in \citet{Natarajan1997}.)
Because the numerical hydrodynamic simulation also accounts for
a variable temperature set by radiative equilibrium, the comparison
in figure 1
also shows that the
departures from isothermal owing to radiative equilibrium
have little effect on the overall
density structure of the core.

Figure \ref{fig:densities} also shows the
characteristic radius, $r_f$, for the different evolutionary times
during the contraction.  The figure shows that equation \ref{eq:criticalRadius}
provides a good approximation to the turn-over radius, $r_f$, at all times
in the evolution. According to equation 2, this width shrinks
as the central density of a
contracting BE sphere increases.
In the subsonic phase of gravitational contraction,
$r_f$ moves at about half the maximum gas velocity where this
maximum is taken from the velocity profiles for each evolutionary
time as shown in figure \ref{fig:velocities}.

In comparing our
theoretical results to several observations, we find that a central
density of about
$2 \times 10^7$ cm$^{-3}$ provides a better match to
the observations (\S \ref{lines}) than central
densities that are
a factor of 10 higher or lower.
This estimate is in agreement with the density of
$2 \times 10^6$ cm$^{-3}$ suggested by the observations
of \citet{Crapsi2007} if we average
the theoretical density profile over the size of their observing
beam.
Although the theoretical density profile
appears flat in the center of a log-log plot of
density versus radius, it is sharply peaked within the
size of the observing beam.
The average density
within a radius of 500 AU
(observing beam size of 1000 AU = 7" at 140 pc)
is $2.8\times 10^6$
corresponding to a peak density of
$2\times 10^7$ cm$^{-3}$.
In fact, at a spatial resolution of 1000 AU, observations
have difficulty measuring the central density. For example, the
average density over 1000 AU corresponding to a peak density of
$2\times 10^8$ cm$^{-3}$ is only $3.4\times 10^6$ cm$^{-3}$ , little
different from the average density of $2.8\times 10^6$ cm$^{-3}$
corresponding to a peak density of $2\times 10^7$ cm$^{-3}$.

In our previous paper \citep{KetoCaselli2008}, we used the lower central
density of $2 \times 10^6$ cm$^{-3}$ and stated that a higher
central density would imply a higher density throughout the core.
This is not correct. Owing to the inside-out
character of the collapse, the central density increases
much more rapidly than the densities outside of $r_f$. As explained
in the next section, in an
evolving core, the densities over most of the core are actually about the
same for both these higher and lower central
densities (figure \ref{fig:densities}).

\subsubsection{The self-similar outer region}

In their numerical hydrodynamic simulations of gravitationally
collapsing gas clouds, \citet{BS1968}  found that regardless
of the initial configuration of the cloud or the initial
conditions, all their examples evolved to density profiles
scaling as $r^{-2}$  in their outer regions.
\citet{Shu1977} suggests that this scaling is a property
shared by all self-gravitating isothermal systems under
certain general conditions. The system should
evolve subsonically, meaning that it should be close to
hydrostatic equilibrium initially, and the outer region
should not be affected by an outer bounding surface. These
conditions are applicable to contracting BE spheres. The
initial state is one of
hydrostatic balance, albeit unstable.  Although
BE spheres have an outer boundary which is the
radius at which the Lane-Emden equation is truncated,
the external pressure at the boundary is the same
as would be provided by the Lane-Emden equation
if the solution were continued. Thus in the initial
configuration, the density profile is the same as if
the core had no outer boundary.

Figure \ref{fig:velocities} shows that the BE sphere
evolves by subsonic contraction to resemble the SIS,
a result also found in earlier studies
\citep{Larson1969, FC1993}.
As the central density
increases with the contraction, the characteristic radius, $r_f$,
moves inward, the constant density region shrinks in size, and
the outer region with its density profile scaling as
$\rho\sim r^{-2}$ accounts for more and more of the core.
Because the density of the outer region changes very slowly,
the density contrast between the inner and outer regions
also increases rapidly. Thus, as the contraction proceeds,
the BE sphere evolves to resemble an SIS with its $r^{-2}$ density profile
everywhere and its infinite density contrast.

This behavior is expected if we recall that
the static equilibrium solutions of the Lane-Emden equation
form a continuous series with increasing density
concentration and the SIS as the limiting case
\citep{Chandrasekhar1957, Shu1977}.
During the subsonic, quasi-static phase of contraction, an unstable
BE sphere evolves approximately along this series of static equilibria
toward the limiting case of an SIS.

\subsection{The velocity profile and the inside-out collapse of a BE sphere}\label{insideout}

The velocity profile during the subsonic phase of the
contraction of a BE sphere (figure 2)
shows a characteristic $\Lambda$-shape with the
inward velocity as a function of radius increasing
from near zero in the core center
to a maximum just outside the characteristic radius, $r_f$,
before decreasing again toward
the edge of the core \citep{FC1993, Ogino1999, Kandori2005}.
The origin of this profile can
be understood by considering the forces
inside and outside of $r_f$. In the very center, the velocity is zero by
symmetry.
In the region, $r_f$, where the density is constant,
the instantaneous acceleration due to the gravitational
force increases linearly with radius. The velocity, which is the
time-integrated acceleration, may also be expected to
increase with radius as well, and this is verified by the
numerical evolution. Outside of $r_f$, the density
falls off as $r^{-2}$. Here the instantaneous gravitational acceleration
decreases with radius along with its influence on the velocity.
During the contraction, the redistribution of mass in the center
does not change the gravitational acceleration in the
outer region because of spherical symmetry. Thus, in
the outer region, the
only change to the hydrostatic balance is due to the
pressure force which propagates outward from the
center only at the sound speed.
These several effects that would change the hydrostatic balance in
the outer region all decrease
as we move outward, away from $r_f$.  Thus the outer
region remains in approximate hydrostatic balance with
low inward velocities for about a sound-crossing time, on the order of 1 Myr,
and the velocities are lower at larger radii, further from $r_f$.

The same characteristic $\Lambda$-shaped profile also develops if we
consider the pressure-free collapse of a BE sphere. In this case the
gas pressure is ignored and the velocity field is determined solely
by the variation of the
gravitational force with radius.
\citep{WhitworthWT2001, Myers2005}. Of course, the gas pressure is
required to obtain the initial BE density distribution in the first place.

\section{The temperature and chemistry of the evolving core}

In the previous section we saw that the dynamics of a contracting
BE sphere, as followed by a numerical simulation that included
radiative and chemical equilibrium, were well matched by
a simpler isothermal approximation. In this section we will
see that the temperature and chemistry of an evolving
core have a significant effect on the observable molecular line spectra.
The gas temperature directly affects the line brightness
through the source function.
The chemistry affects the line brightness through the gas phase abundance
of the observable molecules.
The most significant processes affecting the abundances are
the depletion of molecules from the gas phase as they freeze onto
dust grains in the dense center of a core and photodissociation
of molecules near the core boundary. We calculate the CO
abundance with a simple model that includes these two
processes \citep{KetoCaselli2008}.

Figure \ref{fig:structure123}
shows the gas temperatures, density, velocity,
and CO abundance at 3 different times during the collapse
when the central density is  $2 \times 10^6$, $2 \times 10^7$,
and $2 \times 10^8$ cm$^{-3}$.
As discussed further in \S \ref{lines} , a model with a
central density of $2 \times 10^7$ cm$^{-3}$
provides a better match to the observations than
central densities that are a factor of 10 lower or higher.
Also shown
are the density, temperature,
and abundance of static cores that have the same central densities
as the dynamically evolving core. The properties of the
contracting cores are similar to those of the static cores with
the same central density, with the exception of the velocity.

\subsection{The temperature in the center of a contracting core}

The starless cores are heated from the outside by both cosmic rays and
by the diffuse
interstellar radiation field supplied by  the population of stars in the Galaxy.
The cores are cooled by the long wavelength radiation from cold dust
and by molecular lines.
Because the incoming shorter wavelength starlight is strongly absorbed by the cores
whereas the outgoing long wavelength
radiation from cold dust is not, the denser cores
are coldest in their centers. This temperature structure
has previously been calculated in quite a number of papers
\citep[references in][]{KetoField2005}.  In figure \ref{fig:structure123}
we see that the central temperature of the static core with
a central density of $2 \times 10^8$ cm$^{-3}$ is quite low,
below 5 K. In the evolving core, compressive heating keeps the
temperature above 5 K, more consistent with the temperatures
inferred from molecular line and dust observations 
\citep{Crapsi2007, Pagani2007}.

\subsection{CO depletion and desorption}\label{depletion}

In our previous paper we assumed that the rate for
the desorption of CO off dust, the inverse process to depletion
or freeze-out, was due to 
whole-grain heating by cosmic rays
\citep[equation 11][]{KetoCaselli2008,HasegawaHerbst1993}.
When we compare the C$^{18}$O and C$^{17}$O spectra predicted 
by our radiative transfer
code MOLLIE with observations of L1544 we find that
the predicted lines are not
bright enough to match the observations. Since these
CO lines are optically thin, 
the CO column density is too low.

The desorption rate cannot be increased by simply increasing
the flux of cosmic rays.
In our earlier study we found that a higher flux
of cosmic rays would cause the gas temperature
at mid-radii to be higher than suggested by observations
\citep[figure 12][]{KetoCaselli2008}.

There are other processes in addition to direct cosmic-ray
strikes that cause desorption and increase the gas phase 
abundance of CO \citep{PrasadTarafdar1983,Leger1985,
dHendecourt1985,DuleyWilliams1993,WillacyMillar1998,
TakahashiWilliams2000,Dalgarno2006,Oberg2009}.
\citet{Shen2004} found that the energy transferred to dust by the UV
radiation field produced by cosmic-ray strikes on molecular
hydrogen is almost one order of 
magnitude larger than the energy transferred directly to dust by 
the cosmic-ray particles themselves.
\citet{Roberts2007} suggested that in addition to
direct and indirect heating by cosmic rays, another
significant source of heat is
the exothermic
formation of H$_2$ on the grain surfaces.
\citet{Roberts2007} suggest that the rates of
these heating processes are not known. However,
because all the processes depend
on the first power of the density, the same as direct
cosmic ray heating, we can account for additional
desorption processes in our model by simply increasing
the desorption rate above that given by
equation 11 of \citet{KetoCaselli2008} while still
assuming the same first-power dependence on the gas
density. 

Figure \ref{fig:abundance} shows the abundance
of CO obtained as an equilibrium between
the 4 processes of depletion, desorption, photodissociation
and chemical formation, and calculated for 4 different 
desorption rates.  The lowest rate is
equivalent to desorption caused by direct cosmic-ray
strikes \citep{HasegawaHerbst1993} on dust at
the "low" cosmic-ray rate ($1.3\times 10^{-17}$ s$^{-1}$)
as defined in equation 11 of \citet{KetoCaselli2008}. 
The higher desorption rates are factors of
3, 9, and 30 higher than the lowest rate.
In order to match the observed CO spectra
(figure \ref{fig:spectrum123})
we find that we need to increase our desorption
rate by a factor of 30.  
At this rate, desorption
and depletion have equal time scales at a density of
about $10^4$ cm$^{-3}$.

As shown in
figure \ref{fig:abundance},
with the highest desorption rate, the maximum
CO abundance is about a factor of 2.5 higher
than with the lowest rate.
This increase in abundance makes
the optically thin C$^{17}$O
and C$^{18}$O lines almost a factor of 2 brighter
and a better match to the observations. 

The gas phase abundance could also be increased by decreasing
the reverse reaction of freeze-out. One way this could occur is if
the dust grains in the center of L1544 were fewer in number but
larger in size, for example by coagulation. This would
decrease the total surface area available for freeze-out 
\citep{VanDishoeck93}.  There is some independent evidence for
coagulated grains \citep{Caselli2002,Keto2004,Bergin2006,Flower2006,
VanWeeren2009}. However, to reduce the rate of freeze-out by a factor of 30, 
would require a mean grain diameter of about 10~$\mu$m, too large according 
to available models of dust coagulation in 
dense clouds \citep{OssenkopfHenning1994}. 
Decreased depletion at a rate slower than equation 8 of \citet{KetoCaselli2008} 
could contribute to the higher CO
abundance, but could not bring the CO abundance up high enough
to match the spectral line observations. 
Finally, 
decreased photodissociation might also increase the CO abundance.
This could come about if the UV flux were diminished, for example,
if the core were embedded in a larger scale molecular cloud.
We ran a separate radiative transfer calculation and found
that the CO abundance cannot be increased deep enough
into the core to make a difference to the line brightness.

\subsection{Depletion time scale}

The time scale for CO to freeze onto dust, $\tau_{on}$ 
\citep[equation 8][]{KetoCaselli2008,Rawlings1992}
and the time scale for the
inverse process of desorption, $\tau_{off}$ 
\citep[$30 \times$ equation 11][]{KetoCaselli2008,HasegawaHerbst1993}
may be combined to determine the time scale for the change of CO in
the gas phase \citep{Caselli2002},
\begin{equation}
\tau_{CO} = {{\tau_{on}\tau_{off}}\over{\tau_{on} + \tau_{off}}}.
\label{eq:depletion}
\end{equation}

This time scale for the equilibration of the CO abundance
varies across the core  but is everywhere 
aster than the dynamical time. 
In the core center, the
free-fall time is a close approximation whereas in the more slowly
evolving outer region, the sound crossing time is more appropriate.
A starless core with a mass of 10 M$_\odot$ and a central
density of $2\times 10^6$ cm$^{-3}$ has a free-fall time,
$\tau_{ff}=0.03$ Myr
using the central density in the standard equation whereas the sound
crossing time is about 2 Myr.
In comparison, at the center of the core, the CO equilibration
time scale $\tau_{CO} = 0.01$ Myr, a
factor of 3 shorter than the free-fall time.  At the edge of the core
where the density
is 20 cm$^{-3}$, $\tau_{CO} = 0.5$ Myr, still a factor of 4
shorter than the sound
crossing time. 
throughout the contraction.
Figure \ref{fig:structure123}
shows that the abundance of CO during the contraction
of a BE sphere is everywhere quite close to its steady state value. 
For this reason, it is not possible to use CO as a chemical
clock in starless cores.

\subsection{CO abundance and time reversibility}

The previous discussion demonstrates the importance
of desorption in determining the CO abundance. 
The desorption rate is required to determine
the equilibrium abundance toward which the system evolves asymptotically.
In particular, the CO abundance should not be
calculated simply from the depletion rate
as an exponentially decreasing function of time as would be the case if 
desorption were ignored. 
Even in gas dense enough that the 
depletion rate is much faster,
general considerations always require  both the
forward and reverse reactions. Real physical systems must
satisfy detailed balance and time reversibility \citep{Onsager1931}.
A system described by a single rate equation, for
example depletion only without desorption, does not
satisfy these conditions, cannot describe a physical
system, and this description will produce
misleading results.

\section{Comparison with observations}\label{lines}

All other things being equal,  a collisionally
excited, optically thin line should increase in brightness
with the
path length and the gas
density. Thus molecular lines should be brightest
through the center of the core. Observations of starless cores do
not always show this behavior because of varying molecular
abundances within the core \citep{Tafalla2002}.
For example, figure 2
of \citet{Caselli1999} reproduced here as figure \ref{fig:caselli99}
shows the brightness
of the (1-0) transitions of C$^{17}$O, C$^{18}$O, and N$_2{\rm H}^+$
at a set of positions across the L1544 core from edge to center to edge.
The N$_2{\rm H}^+$ molecule does not suffer much,
if any, depletion from the gas phase at high densities, and the observed
N$_2{\rm H}^+$ brightness increases toward the core center whereas
the observed CO brightness does not. This difference is due to the near complete
depletion of CO at higher density.
It is possible that N$_2$H$^+$ may
be slightly depleted in the center of L1544 \citep{Caselli2002} although
we do not include this effect in our model. We will investigate this
possibility in a future modeling of higher density tracers and transitions.
Our modeling here shows that
the decrease in temperature in
the core center is not sufficient to significantly
decrease the line brightness.

Figure \ref{fig:trace123}  shows the simulated
integrated spectral line brightness
as a trace of positions across the model evolving cores
in the same format as the observational data in figure \ref{fig:caselli99}.
The integrated intensities include the emission from all three hyperfine
lines of C$^{17}$O(1--0) and all seven hyperfine lines of N$_2$H$^+$(1--0).
At central densities greater than $10^6$ cm$^{-3}$  the modeled brightness
begins to look
like the observations. There are some differences.

The observational data might show a slight dip in the integrated intensity
of the CO lines toward the center which is not present in the model. However,
there is a decrease in the peak intensity of the modeled CO lines
(figure \ref{fig:peak123}).
This dip does not show up in the integrated intensity because the line widths
in the model increase toward the center owing to the increase in the inward 
velocities in the inside-out collapse (figures \ref{fig:width123}).
The increase in the line width of CO is less than that of
N$_2{\rm H}^+$ because, owing to depletion, there is essentially no gas phase CO
in the center of the core where the velocities are highest.
From a density of $2\times 10^6$ to $2\times 10^8$ cm$^{-3}$
the C$^{17}$O and C$^{18}$O line widths increase by 0.07 and 0.06 kms$^{-1}$
while the N$_2$H$^+$ line width increases by 0.15 kms$^{-1}$.

\begin{figure}%[t]
\includegraphics[width=3.25in]{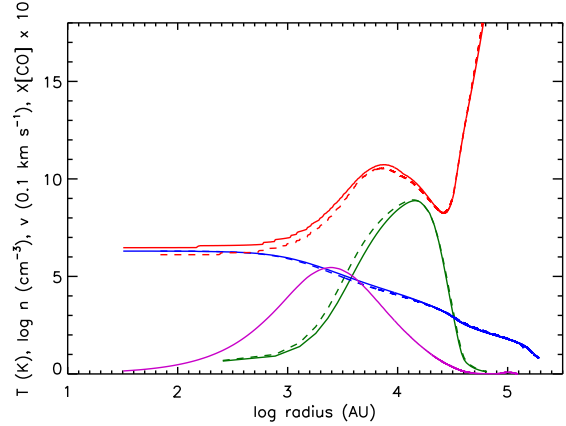}
\includegraphics[width=3.25in]{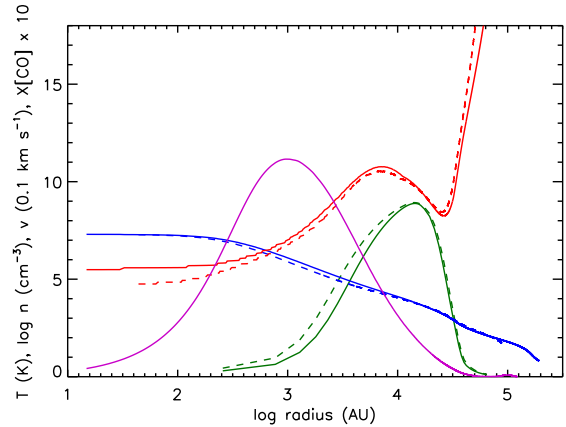}
\includegraphics[width=3.25in]{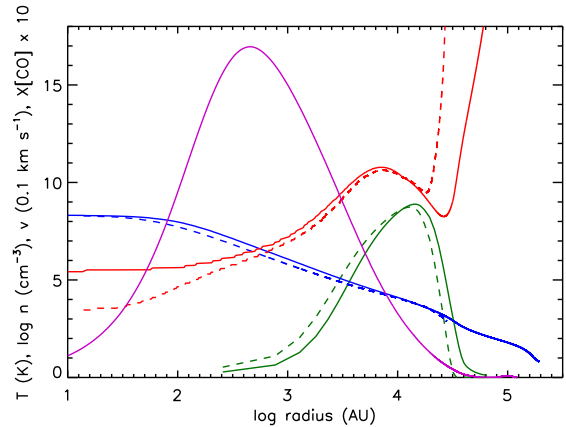}
\caption{The temperature, density, abundance, and inward velocity profiles
of a contracting starless core at 3 times in the evolution of the contraction,
when the central density is $2\times 10^6$ cm$^{-3}$ ({\it top}),
$2\times 10^7$ cm$^{-3}$ ({\it middle}),
and $2\times 10^8$ cm$^{-3}$ ({\it bottom}).
Red indicates the temperature in K, blue the log of the
density in cm$^{-3}$, green
the ratio of the CO abundance to the undepleted abundance multiplied by 10.
A ratio of 10 corresponds
to no depletion
and zero to total depletion.
Purple indicates the inward velocity. In this figure a positive velocity is inward,
and the velocity units are 0.1 kms$^{-1}$.
The dashed lines show the temperature, density, and abundance
of a static sphere with the same central density. The subsonically
contracting sphere closely resembles the static sphere except
for the velocities.
}
\label{fig:structure123}
\end{figure}

\begin{figure}%[t]
\includegraphics[width=3.25in]{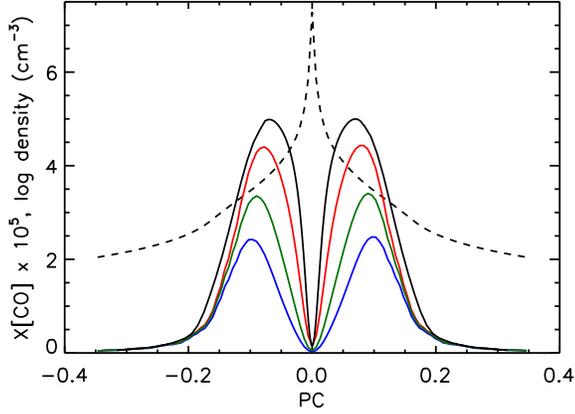}
\caption{The abundance of CO for different rates of desorption.
The dashed black line shows the log
of the density (cm$^{-3}$). The other 4 curves
show the abundance calculated from an equilibrium between
freeze-out, desorption, and photodissociation for 4 different rates of desorption
(\S \ref{depletion}).  The lowest
rate of desorption (blue curve) results in a CO
abundance that is too low to match the observed CO line brightness.
The highest rate of desorption (black curve)
results in spectral line models that
best match the observations.
}
\label{fig:abundance}
\end{figure}

\begin{figure}%[t]
\includegraphics[width=3.25in]{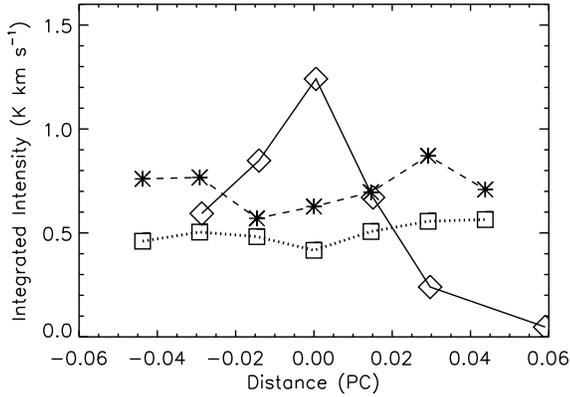}
\caption{Observations of the trace of the integrated intensity of C$^{18}$O(1-0)
(boxes and dotted line), C$^{17}$O(1-0) (stars and dashed line), and N$_2$H$^+$(1-0)
(diamonds and solid line)
across L1544 from \citet{Caselli1999}. The N$_2$H$^+$ molecule is not much
depleted, if at all, at high densities, and its emission peaks toward the center of
L1544. In contrast, the lines of CO, a molecule that is subject to depletion, are
approximately constant across the core, possibly with lower intensity toward
the center. The N$_2$H$^+$ emission has been divided by 5.65 and the
C$^{18}$O emission by 3.65. to match figure 2 in \citet{Caselli1999}}
\label{fig:caselli99}
\end{figure}

\begin{figure}%[t]
\includegraphics[width=3.25in]{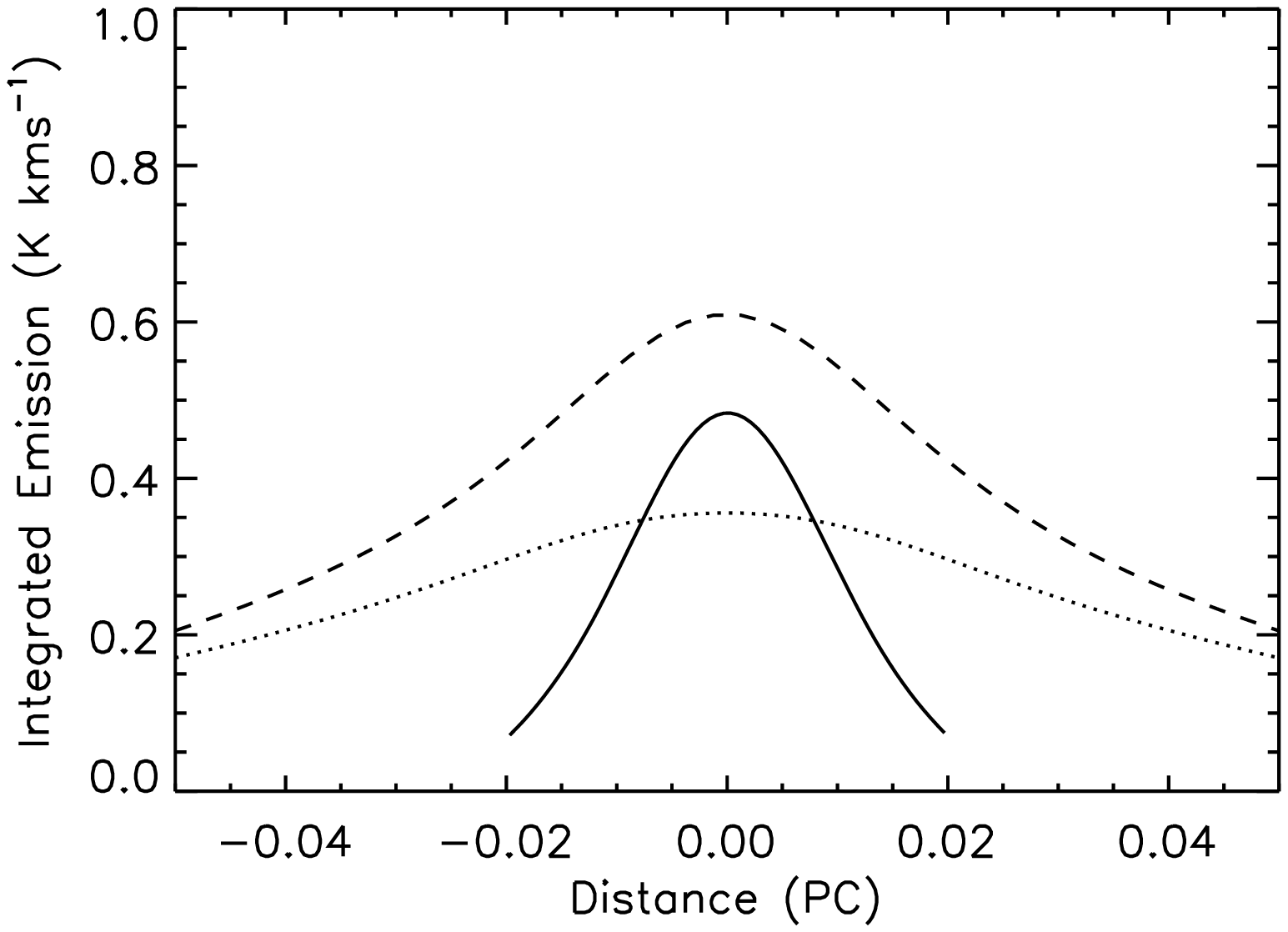}
\includegraphics[width=3.25in]{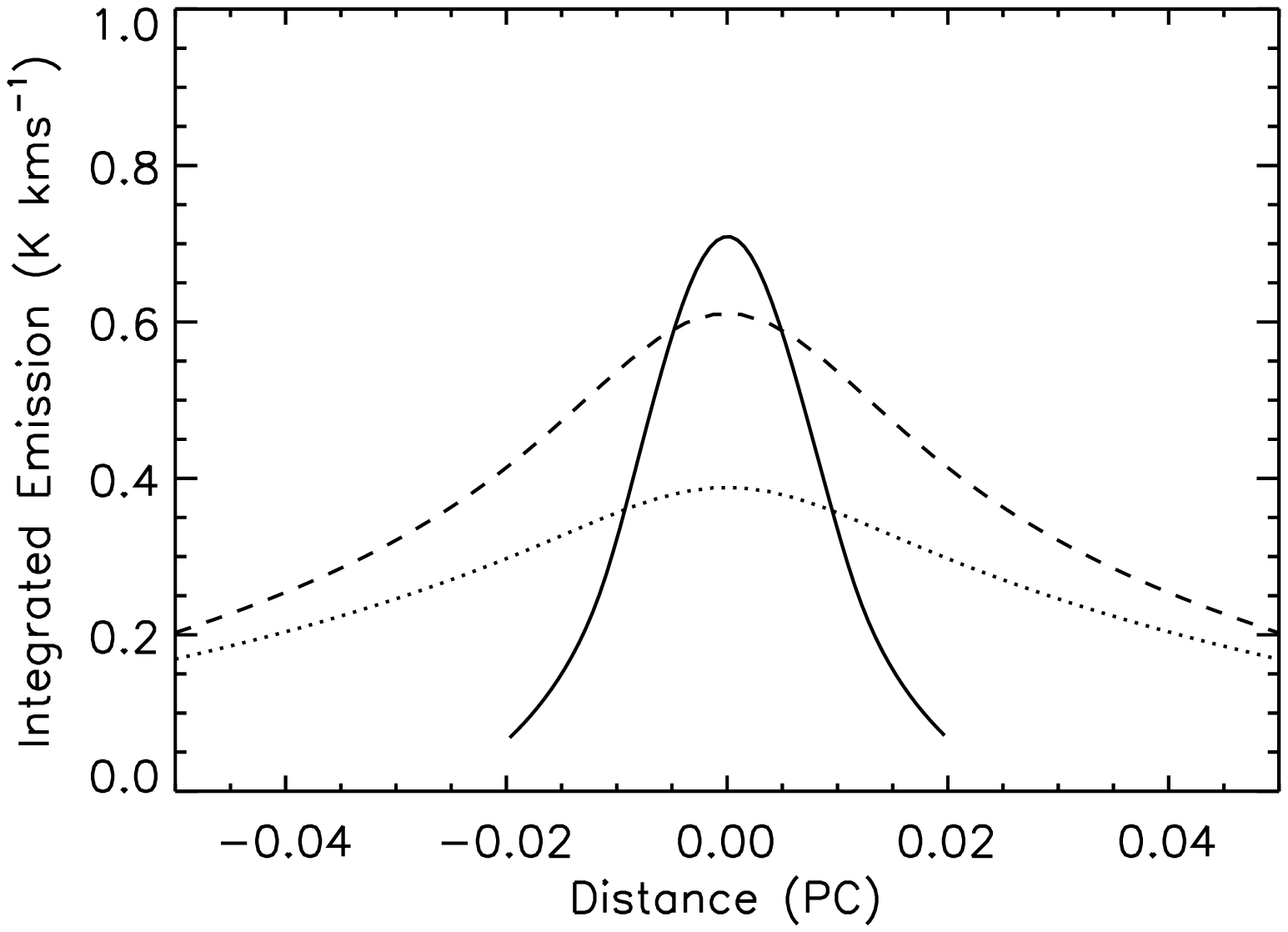}
\includegraphics[width=3.25in]{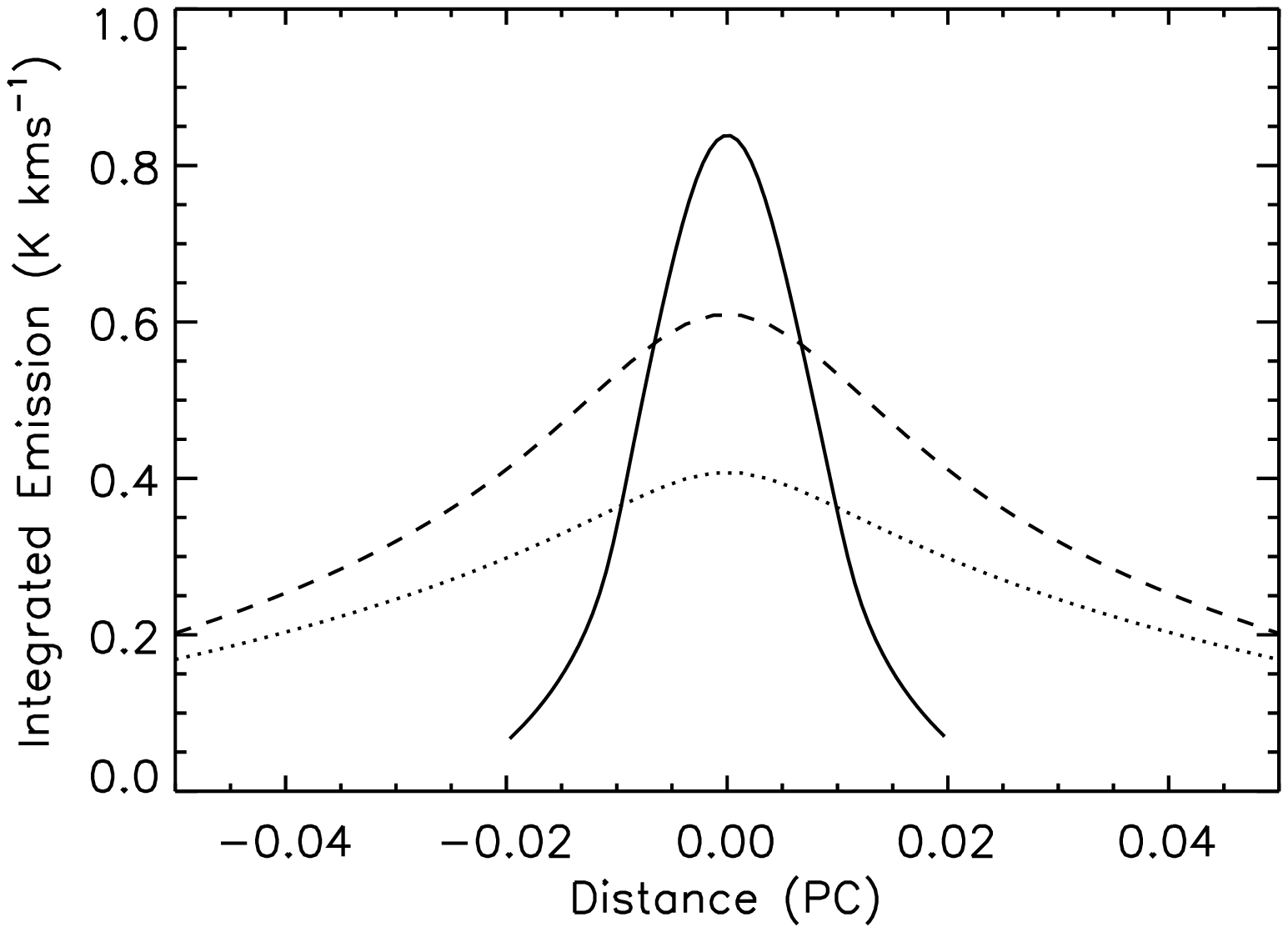}
\caption{Trace of the integrated intensity of  C$^{18}$O(1-0) ({\it dotted line}),
C$^{17}$O(1-0) ({\it dashed line}),
and N$_2$H$^+$(1-0) ({\it solid line}), across our model of a subsonically contracting core.
This figure shows the integrated intensities at 3 times in the evolution of the contraction,
when the central density is $2\times 10^6$ cm$^{-3}$ ({\it top}),
$2\times 10^7$ cm$^{-3}$ ({\it middle}),
and $2\times 10^8$ cm$^{-3}$ ({\it bottom}).
The internal structure of this model is shown in figure \ref{fig:structure123}.
The N$_2$H$^+$ emission has been divided by 5.65 and the
C$^{18}$O emission by 3.65, the same as in figure \ref{fig:caselli99}.}
\label{fig:trace123}
\end{figure}

\begin{figure}%[t]
\includegraphics[width=3.25in]{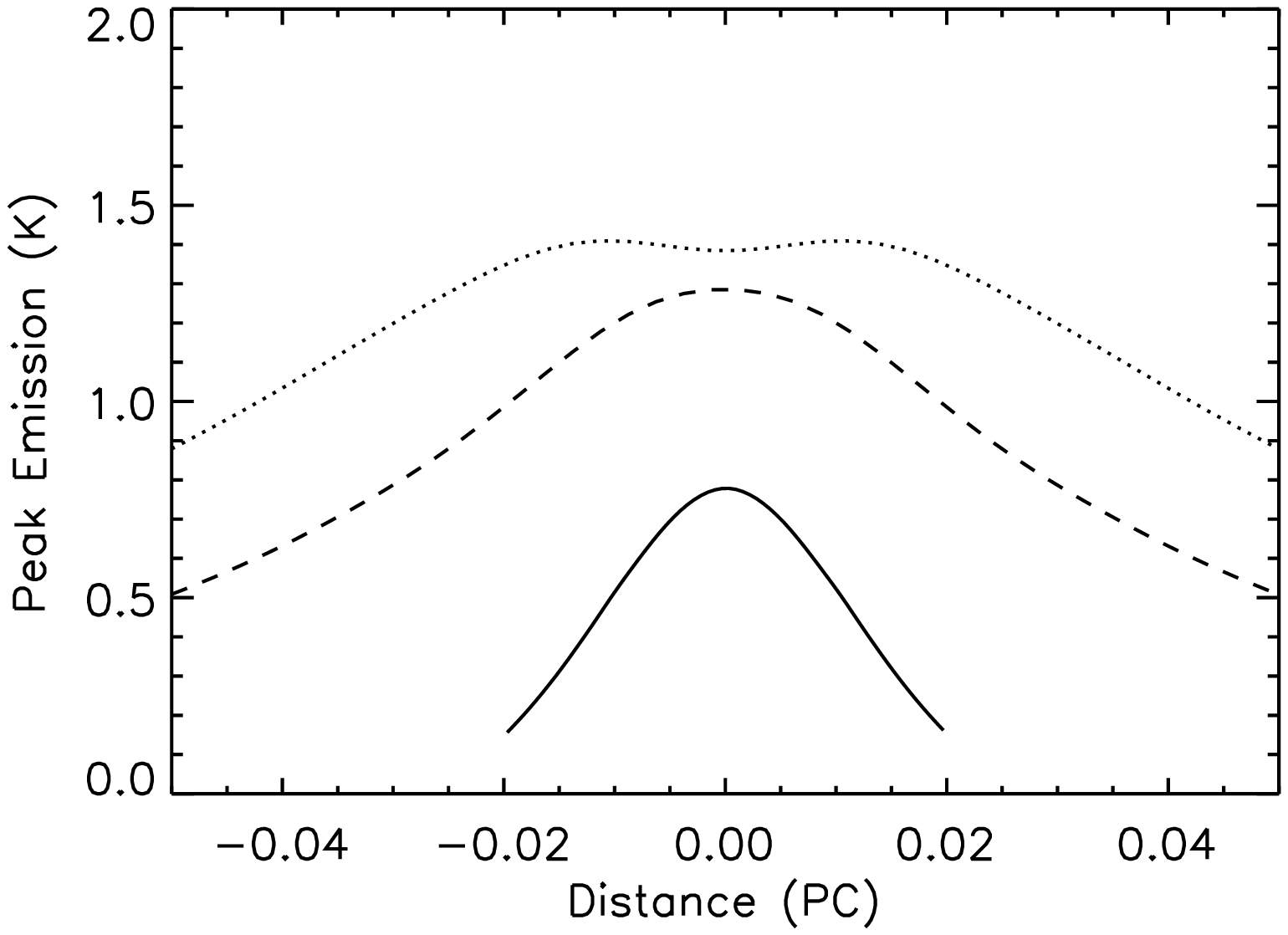}
\includegraphics[width=3.25in]{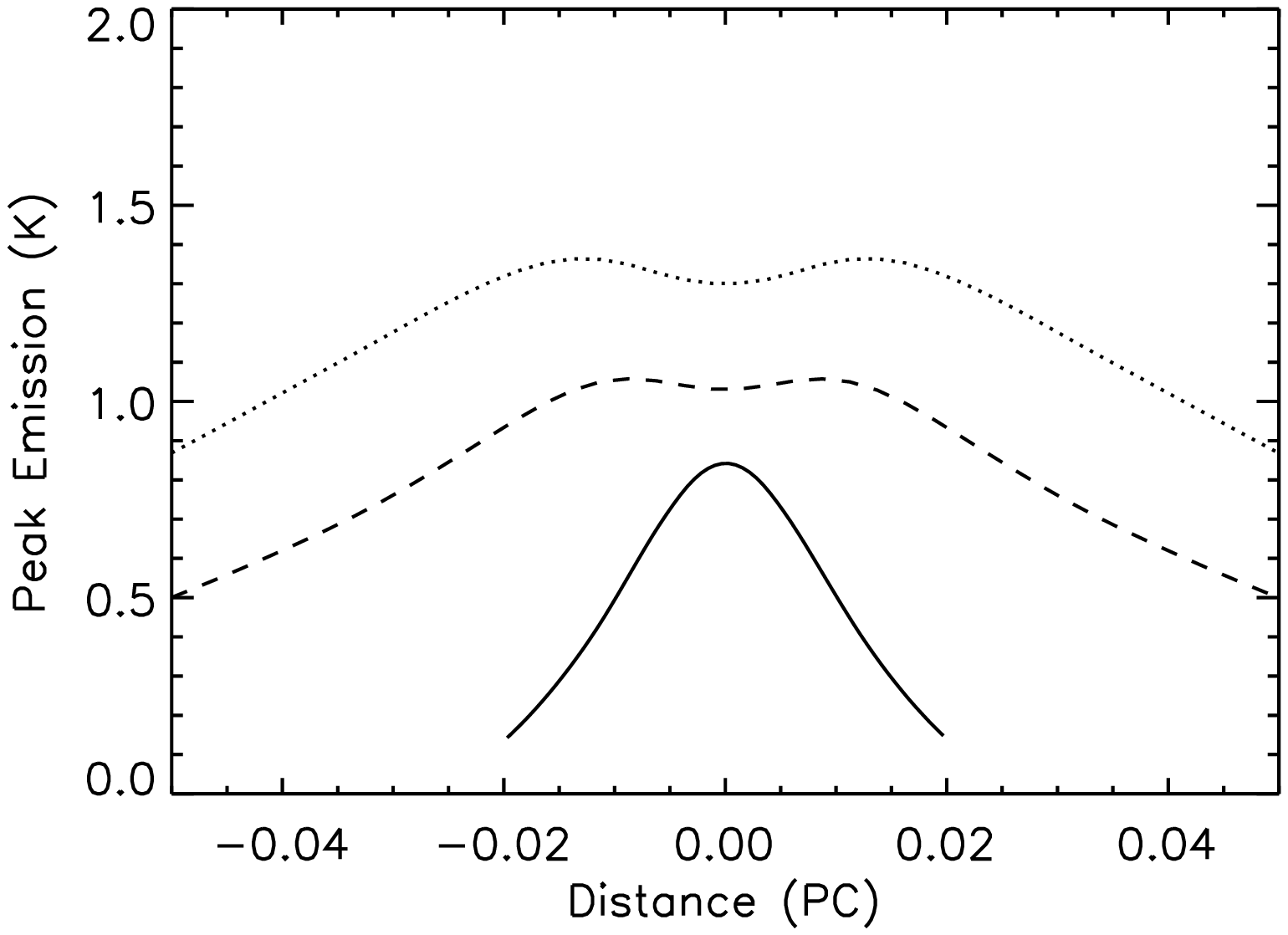}
\includegraphics[width=3.25in]{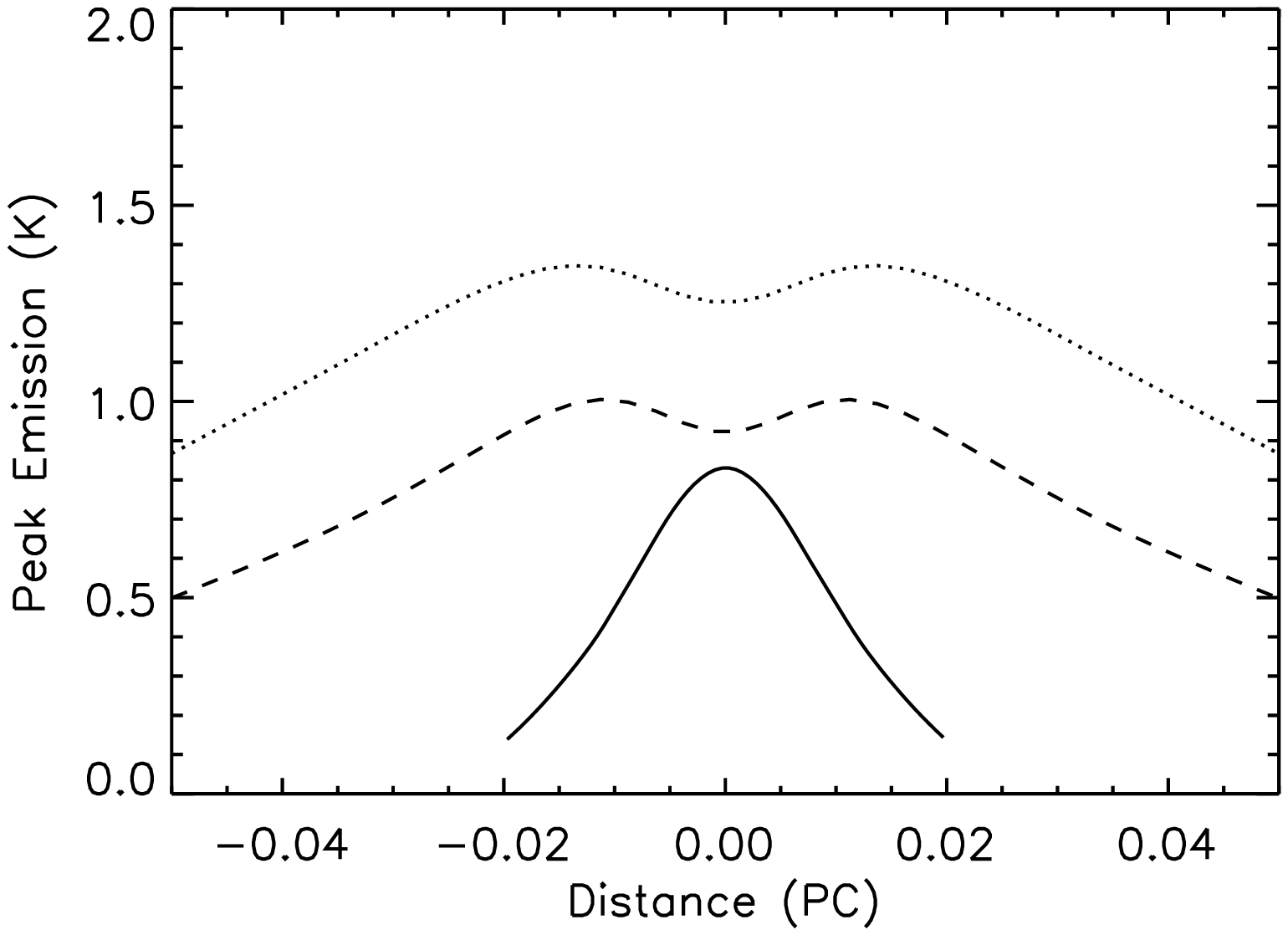}
\caption{Trace of the peak brightness of  C$^{18}$O(1-0) ({\it dotted line}),
C$^{17}$O(1-0) ({\it dashed line}),
and N$_2$H$^+$(1-0) ({\it solid line}),
across our model of a subsonically contracting core.
This figure shows the peak brightness at 3 times in the evolution of the contraction,
when the central density is $2\times 10^6$ cm$^{-3}$ ({\it top}),
$2\times 10^7$ cm$^{-3}$ ({\it middle}),
and $2\times 10^8$ cm$^{-3}$ ({\it bottom}).
The internal structure of this model is shown in figure \ref{fig:structure123}.
The N$_2$H$^+$ emission has been divided by 5.65 and the
C$^{18}$O emission by 3.65, the same as in figure \ref{fig:caselli99}.
The ratio of the peak intensity of C$^{18}$O(1-0) and C$^{17}$O(1-0)
is different than the ratios of the integrated intensities because the
C$^{17}$O(1-0) line is split into 3 hyperfine lines (figure \ref{fig:spectrum123}).
}
\label{fig:peak123}
\end{figure}

\begin{figure}%[t]
\includegraphics[width=3.25in]{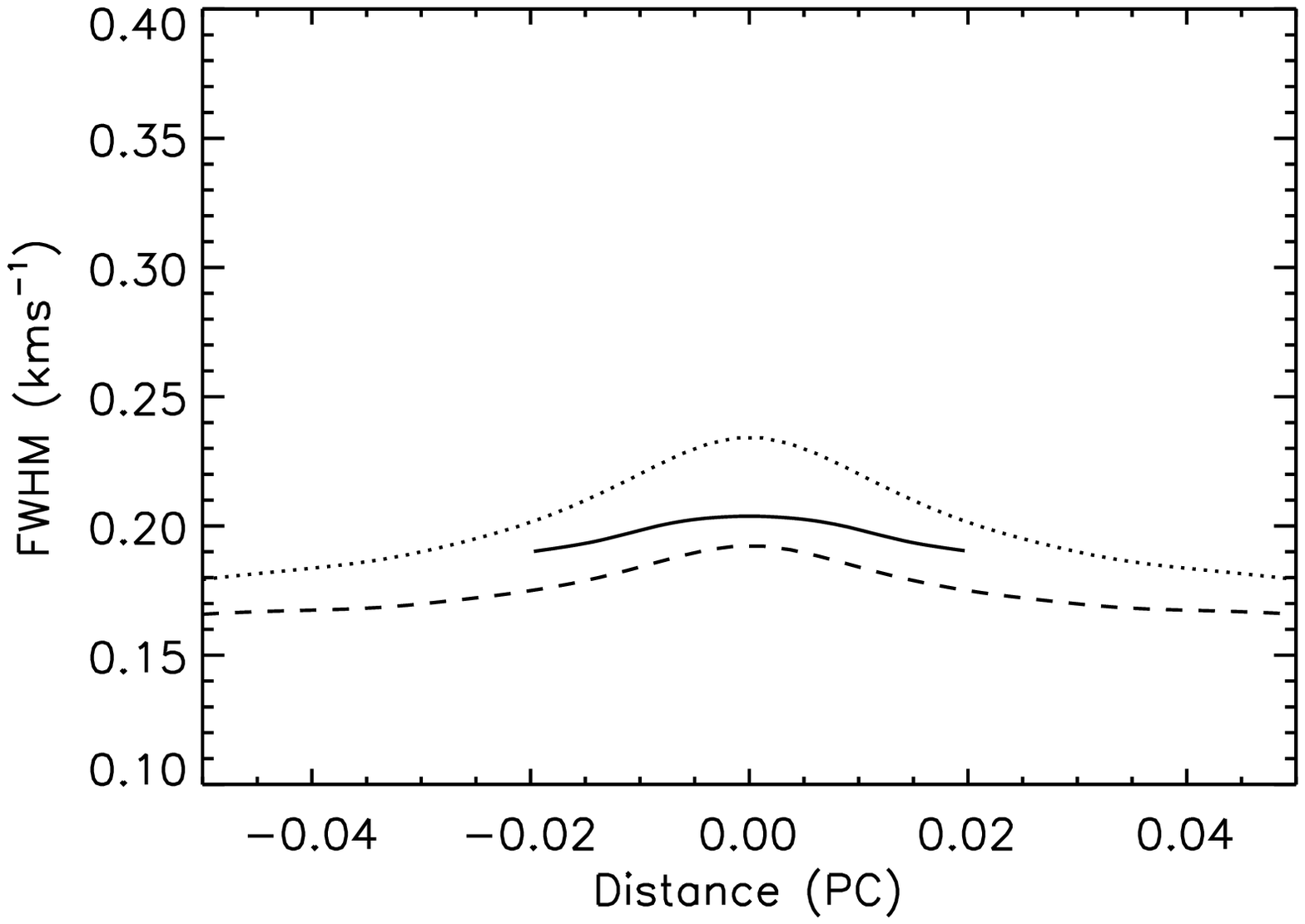}
\includegraphics[width=3.25in]{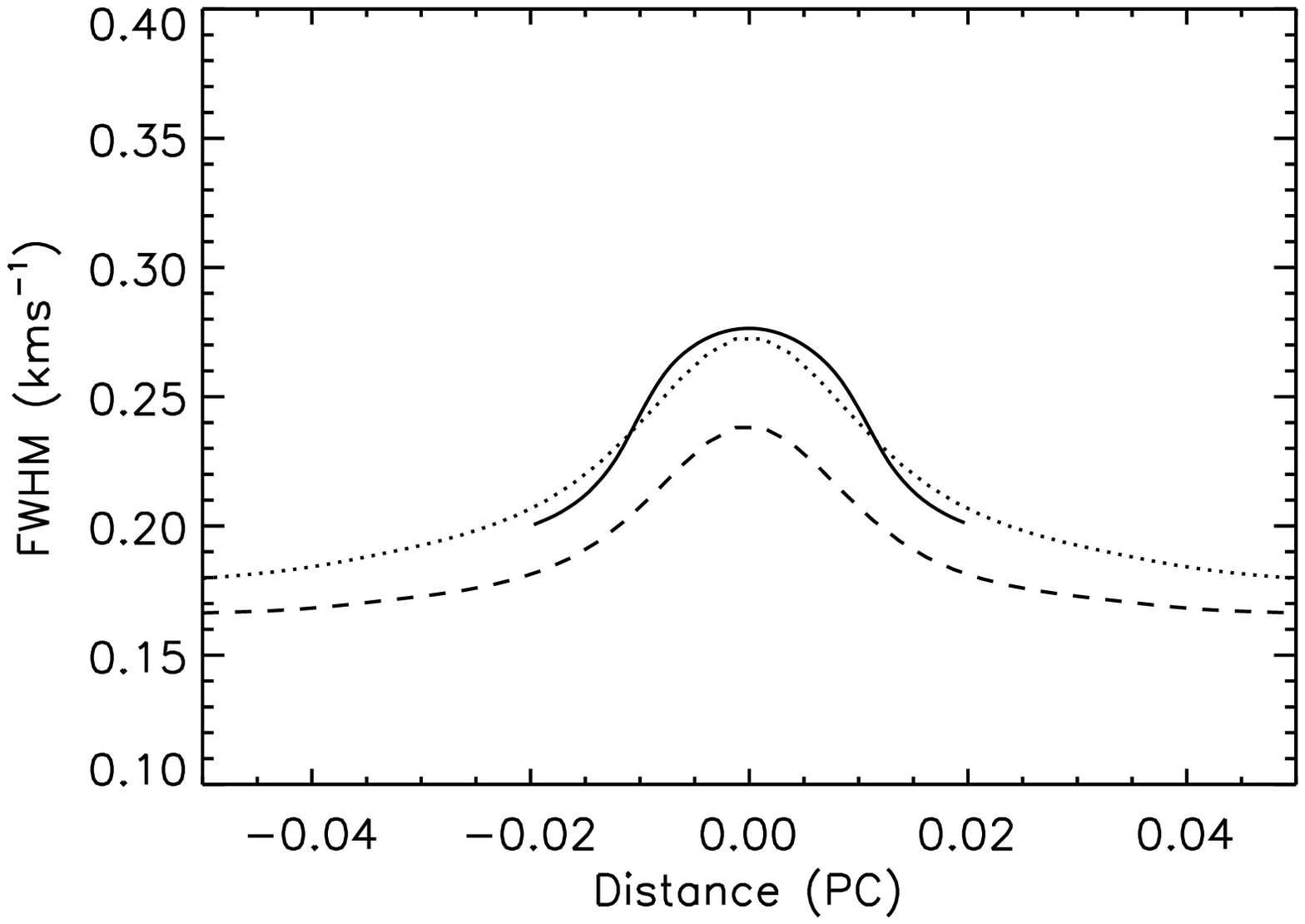}
\includegraphics[width=3.25in]{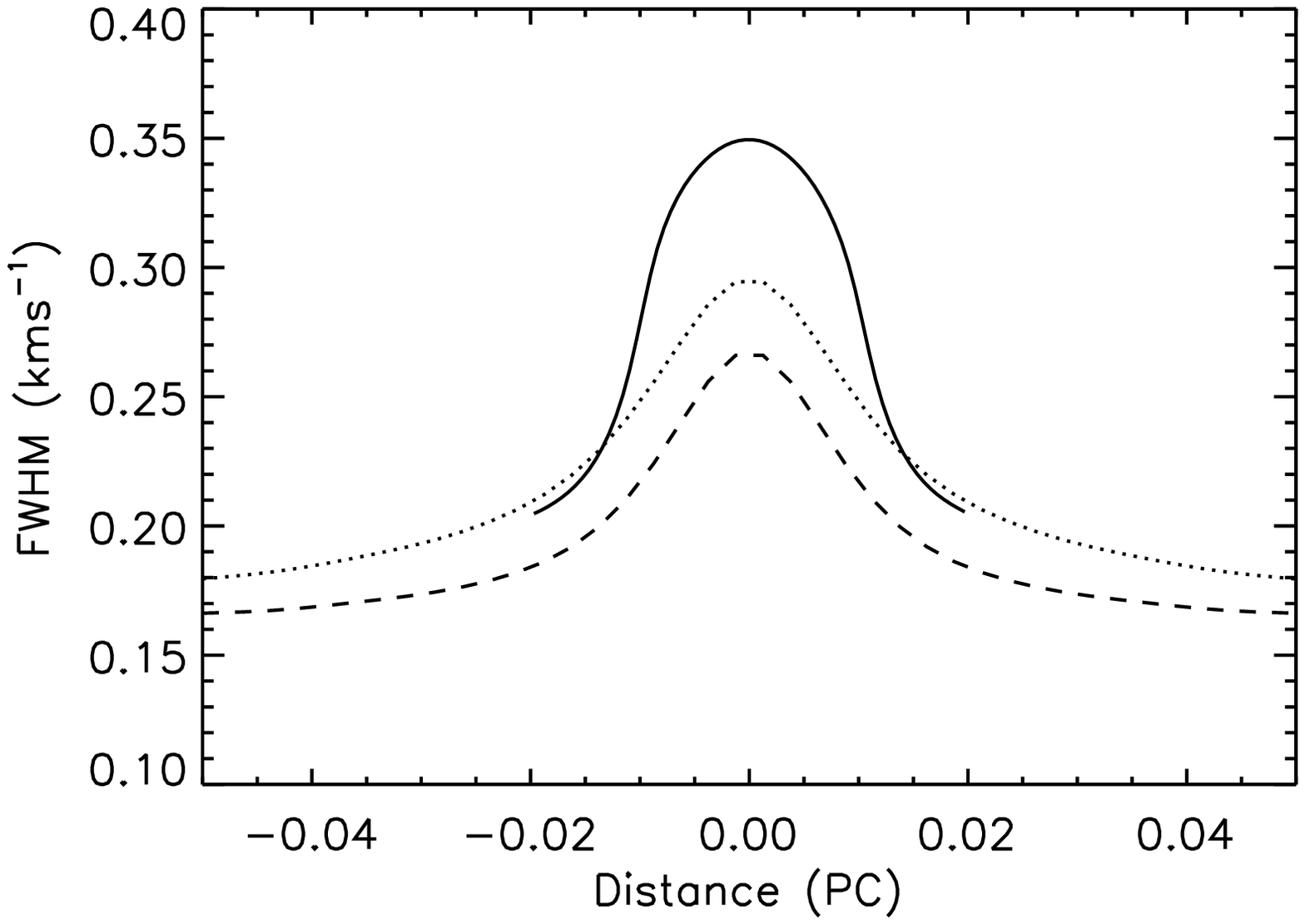}
\caption{Trace of the FWHM of  C$^{18}$O(1-0) ({\it dotted line}),
C$^{17}$O(1-0) ({\it dashed line}),
and N$_2$H$^+$(1-0) ({\it solid line}), across our model of a subsonically contracting core.
This figure shows the line width at 3 times in the evolution of the contraction,
when the central density is $2\times 10^6$ cm$^{-3}$ ({\it top}),
$2\times 10^7$ cm$^{-3}$ ({\it middle}),
and $2\times 10^8$ cm$^{-3}$ ({\it bottom}).
The figure shows that in the center of the core the predicted
C$^{18}$O, C$^{17}$O, and N$_2$H$^+$ line widths increase by 0.074, 0.060, and 0.145 kms$^{-1}$,
respectively because the unresolved infall velocities
increase as the core evolves.
The internal structure of this model is shown in figure \ref{fig:structure123}.
}
\label{fig:width123}
\end{figure}

Because the density, temperature, and abundance
structures of contracting cores are so similar to those of
static cores, the primary observable evidence of contraction
is not the line brightness, but rather
the shapes of those spectral lines such as N$_2{\rm H}^+$(1-0) that are
optically thick enough to show self-absorption.
Figure \ref{fig:spectrum123}  shows the
spectral line profiles of the (1-0) transitions of
C$^{17}$O, C$^{18}$O, and N$_2{\rm H}^+$
expected from our model for the 3 evolutionary times corresponding
to figure \ref{fig:structure123}. As the inward velocities increase
during the contraction, the N$_2$H$^+$
spectral lines become wider (figure \ref{fig:width123}), and
eventually split into two peaks (figure \ref{fig:spectrum123})
because of the
inward velocities in the core (figure \ref{fig:structure123}).
In contrast, because there is little
CO in the center where the velocities are highest,
the width of the CO lines changes very little (figure \ref{fig:width123})
as the core evolves, and the CO lines do not split
(figure \ref{fig:spectrum123}).
The brightness of 
both the N$_2$H$^+$ and CO lines are nearly
constant as the central density increases from $2\times 10^6$
to $2\times 10^8$ cm$^{-3}$ because there is very little
change in the density of most of the core except for
the center where CO is depleted. The slight decrease in the
peak brightness of C$^{18}$O
(figure \ref{fig:peak123} and \ref{fig:spectrum123})
is consistent with a constant integrated intensity and
the slight increase in line width  (figure \ref{fig:width123}). 
As the core evolves from
a central density of $2\times 10^7$ to $2\times 10^8$ cm$^{-3}$
there is
very little change in any of the observed spectra, either CO or N$_2$H$^+$,
because the density increase is happening in a very small region in the
center that does not contain much mass relative to the entire core
and is also becoming small compared to the observing beam.
Thus the N$_2$H$^+$ spectra
(figure \ref{fig:spectrum123})
for the last two evolutionary times look almost the same.
Interferometric observations of higher density transitions of N$_2$H$^+$
and N$_2$D$^+$ will be needed to unveil the dynamical stage, structure and 
kinematics of the rapidly contracting center of the core. This will be 
investigated in a future paper.

\section{Conclusions}

We follow the contraction of a thermally supercritical core through
the evolutionary phase of subsonic contraction with a numerical code that
includes radiative and chemical equilibrium.

We identify a characteristic radius, the product of the sound speed
and the free-fall time, as the point where the density profile of a BE sphere
transitions from an outer region where the density scales as $r^{-2}$
to an inner region of constant or flat density. In the inner region, the
dynamical time is the gravitational free-fall time at the central
density. In the outer region, the dynamical time is the much slower
sound-crossing time. The difference results in inside-out collapse.

Once this characteristic radius becomes smaller than the angular
resolution, observations have difficulty in determining the exact
central density even if the observed molecule is not significantly
depleted.

We follow the gas temperature and abundance of CO during contraction.
In the rapidly contracting center of the core, compressive heating raises
the gas temperature by a few degrees over its value in static equilibrium
and the results are consistent with observations.  The temperature variations
from the radiative equilibrium do not significantly affect the dynamics of
the large scale structure of the thermally supercritical core. The time
scale for the equilibration of CO is everywhere shorter than the dynamical timescale.
Thus, the CO abundance is always close to that of a static sphere of the same density profile.
Therefore, the CO abundance cannot be used a chemical clock to estimate
the age of a starless core.

The comparison with line profiles observed toward L1544 suggests that the
dust has an opacity consistent with ice covered grains, the cosmic ray
ionization rate is close to 1$\times$10$^{-17}$ s$^{-1}$,
and the maximum 
density is about 2$\times$10$^7$ cm$^{-3}$, equivalent to an average
density of 3$\times$10$^6$ cm$^{-3}$ within a radius of 500 AU corresponding
to an observing beam size of 1000 AU.
The line width and intensity of C$^{18}$O and C$^{17}$O lines can be
simultaneously reproduced by our model only if the CO desorption rate is
about 30 times larger than the rate expected from cosmic-ray impulsive heating alone.

\bsp

\clearpage

% plot program /sma/ketoSci/L1544/high_desorption/collapse.pro

%\clearpage

% plot program /sma/ketoSci/L1544/high_desorption/collapse.pro,/density

%\vfill
%\clearpage

%plot file: /sma/ketoSci/L1544/high_desorption/diff4b_t1_structure
%   diff4b_t2_structure
%   diff4b_t3_structure

\clearpage

%/sma/ketoSci/L1544/high_desorption/plot_abundance.pro

 % plot file /sma/keto/L1544/high_desorption/caselli99

%\clearpage
%-----------------------------------------------------------------

%plot file /sma/ketoSci/L1544/high_desorption/new_structure4b.pro,/conv,/integrated,model=1,2,3

\clearpage
%-----------------------------------------------------------------
%-----------------------------------------------------------------

%plot file /sma/ketoSci/L1544/high_desorption/new_structure4b.pro,/conv,model=1,2,3

%\clearpage
%-----------------------------------------------------------------
%-----------------------------------------------------------------

%plot file /sma/ketoSci/L1544/high_desorption/new_structure4b.pro,/width,/conv,model=1,2,3

\clearpage
%-----------------------------------------------------------------

\begin{figure*}%[t]
\includegraphics[width=3.25in]{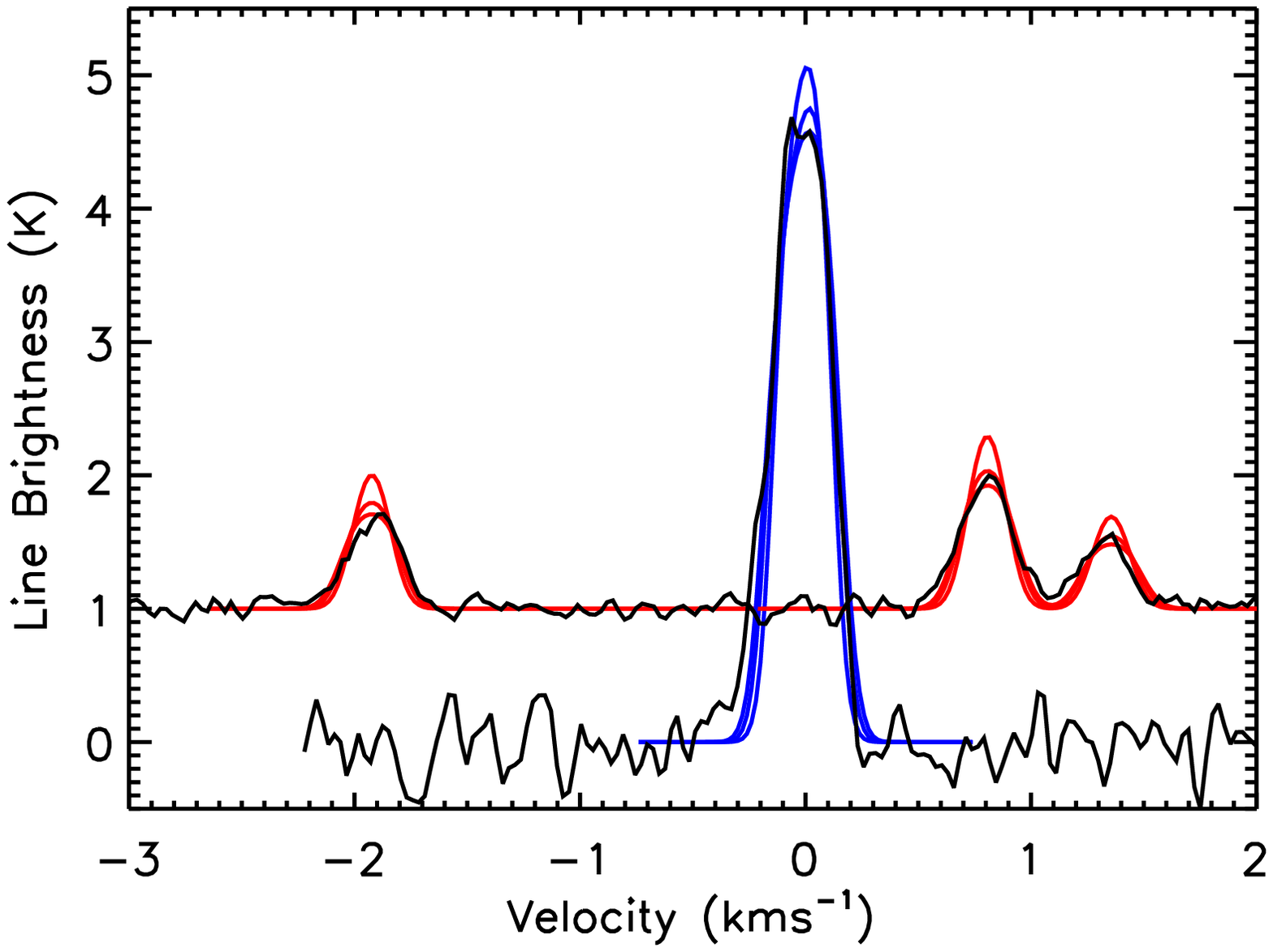}
\includegraphics[width=3.25in]{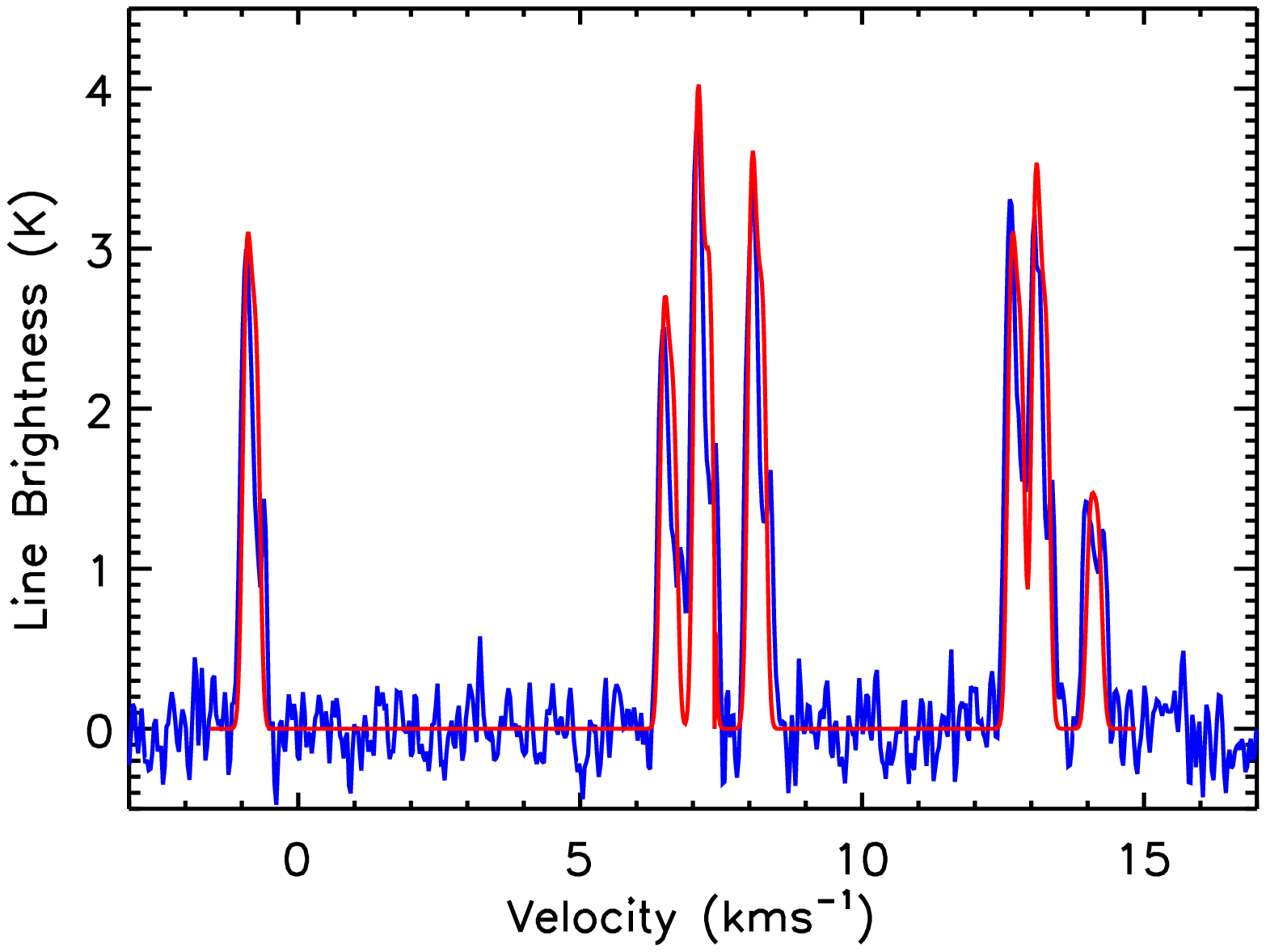} 
\includegraphics[width=3.25in]{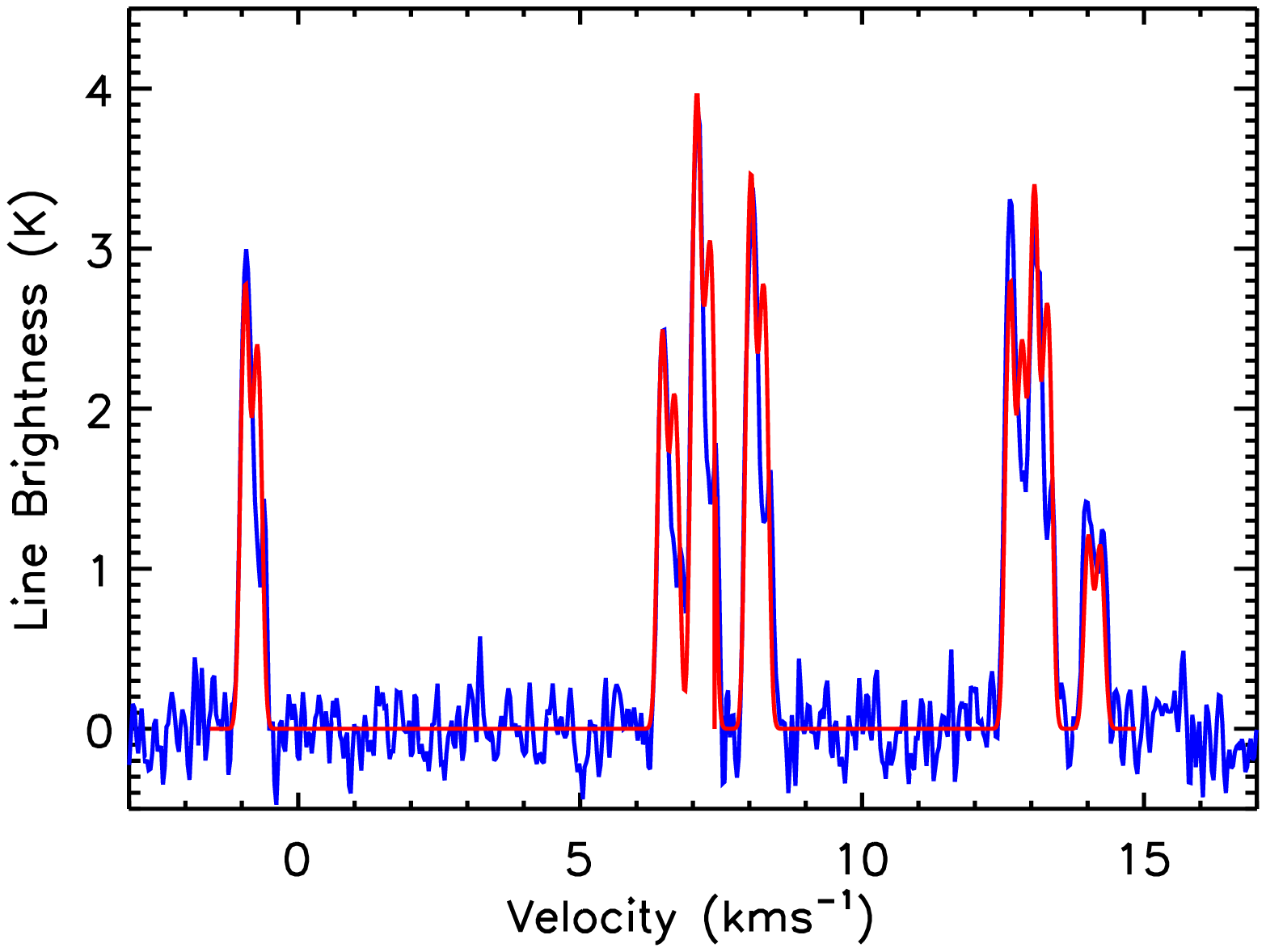}
\includegraphics[width=3.25in]{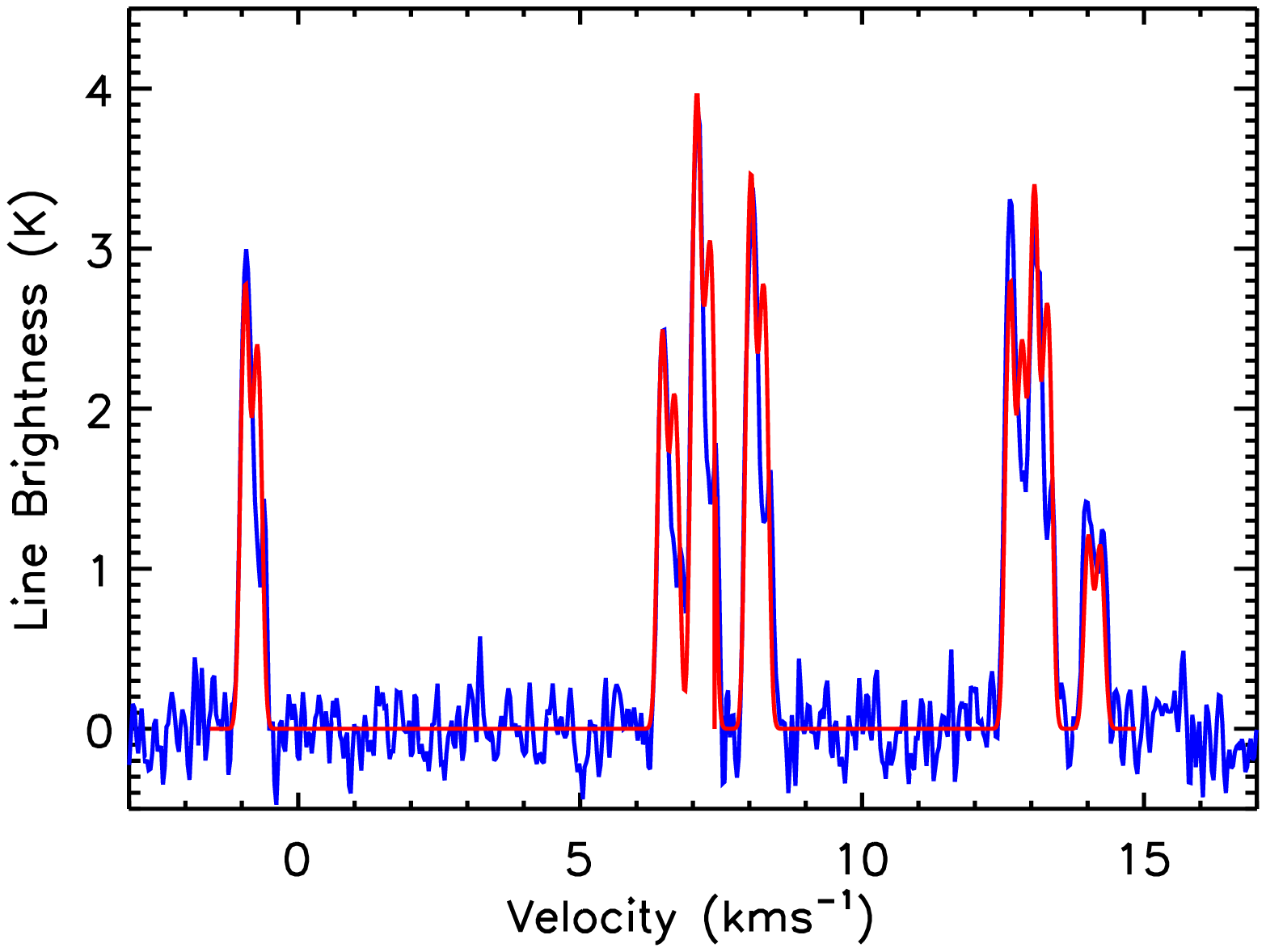}
\end{figure*} 
\begin{figure*}
\caption{Spectra of CO(1-0)
and N$_2$H$^+$(1-0) at the center of our model of a subsonically
contracting core.
This figure shows the line emission at 3 times in the evolution of the contraction,
when the central density is $2\times 10^6$ cm$^{-3}$,
$2\times 10^7$ cm$^{-3}$,
$2\times 10^8$ cm$^{-3}$.
The {\it upper left} figure shows  spectra of
C$^{18}$O(1-0) ({\it red}), C$^{17}$O(1-0) ({\it blue})
at the 3 evolutionary times corresponding to the different densities.
Also shown are the observed spectra \citep{Caselli1999}.
The  C$^{17}$O spectra have been shifted upward by 1 K.
There is very little difference in the model CO spectra as the core evolves.
A slight
decrease in peak intensity is noticeable from $2\times 10^6$ 
to $2\times 10^8$ yr as the line width increases with the increasing infall
velocities.
The other 3 figures show the spectra of N$_2$H$^+$(1-0) at the 3 times
with one figure for each time:
$2\times 10^6$ cm$^{-3}$ ({\it upper right}),
$2\times 10^7$ cm$^{-3}$ ({\it lower left}), and
$2\times 10^8$ cm$^{-3}$ ({\it lower right}). In these 3 figures, each N$_2$H$^+$
model spectrum ({\it red}) is shown together with the observed spectrum ({\it blue}).
There is very little difference between the model spectra at the last two evolutionary
times because the density increase is confined to a small region in the very
center of the core. The internal structure of these models are
shown in figure \ref{fig:structure123}.
}
\label{fig:spectrum123}
\end{figure*}
%plot file /sma/ketoSci2/L1544/high_desorption/co3_spectra.pro
%plot file /sma/ketoSci2/L1544/high_desorption/n2h_spectra.pro,model=1,2,3

\end{document}